\documentclass[conference]{IEEEtran}

\usepackage[a4paper, left=0.680in, right=0.680in, bottom=1.049in, top=0.71in]{geometry}

\usepackage{ifpdf}
\ifCLASSINFOpdf
\usepackage[pdftex]{graphicx}
\graphicspath{{./Figures/}}
\DeclareGraphicsExtensions{.pdf,.jpeg,.png}
\else
\usepackage[dvips]{graphicx}
\DeclareGraphicsExtensions{.pdf}
\usepackage[caption=false, font=footnotesize]{subfig}
\fi

\usepackage{tabularx, booktabs}
\usepackage{amssymb}
\usepackage{amsthm} 
\usepackage[ruled,vlined,linesnumbered]{algorithm2e}
\usepackage{aligned-overset} 
\usepackage{array}
\usepackage{balance}
\usepackage{cite}
\usepackage{color}
\usepackage{epstopdf}
\usepackage{enumitem} 
\usepackage{mathtools} 
\usepackage{makecell} 
\usepackage{multicol} 
\usepackage{multirow} 
\usepackage[caption=false,font=small]{subfig}
\usepackage{textcomp}
\usepackage{stfloats}
\usepackage{url}
\usepackage{verbatim}
\usepackage{graphicx}

\newcommand{\mat}[1]{{\mathbf{#1}}}

\newcommand\blfootnote[1]{%
  \begingroup
  \renewcommand\thefootnote{}%
  \footnote{#1}%
  \addtocounter{footnote}{-1}%
  \endgroup
}

\begin{document}

\title{Temporal Micro-Doppler Spectrogram-based ViT Multiclass Target Classification}

\author{
    \IEEEauthorblockN{Nghia~Thinh~Nguyen~and~Tri~Nhu~Do}
    \IEEEauthorblockA{Telecom Neural Detection Lab, Department of Electrical Engineering, Polytechnique Montr\'{e}al, Montreal, QC, Canada.}
    \IEEEauthorblockA{ Emails: nghia-thinh.nguyen@etud.polymtl.ca, tri-nhu.do@polymtl.ca }
	}
\maketitle

\begin{abstract}
In this paper, we propose a new Temporal MDS-Vision Transformer (T-MDS-ViT) for multiclass target classification using millimeter-wave FMCW radar micro-Doppler spectrograms. Specifically, we design a transformer-based architecture that processes stacked range-velocity-angle (RVA) spatiotemporal tensors via patch embeddings and cross-axis attention mechanisms to explicitly model the sequential nature of MDS data across multiple frames. The T-MDS-ViT exploits mobility-aware constraints in its attention layer correspondences to maintain separability under target overlaps and partial occlusions. Next, we apply an explainable mechanism to examine how the attention layers focus on characteristic high-energy regions of the MDS representations and their effect on class-specific kinematic features. We also demonstrate that our proposed framework is superior to existing CNN-based methods in terms of classification accuracy while achieving better data efficiency and real-time deployability.
\end{abstract}

\begin{IEEEkeywords}
Radar, Classification, MDS, mmWare, FMCW, CNN, ViT, Explainable AI
\end{IEEEkeywords}

\vspace{-.8em}
\section{Introduction}
The millimeter-wave (mmWave) frequency-modulated continuous-wave (FMCW) radar has become an attractive modality for multi-target classification based on micro-Doppler spectrogram (MDS) representations \cite{Zhu2022Continuous}. Recent works show that MDS acquired from one or multiple mmWave nodes can support reliable recognition of human activities and dynamic object states in realistic environments \cite{Zhu2022Continuous,Kim2022RTD}. In multiclass target scenarios, however, superposed returns and partial occlusions cause MDS to overlap in the time–frequency–angle space, so that signatures from different movers interfere and the separability of targets varies with their trajectories, aspect angles, and speeds \cite{Zhu2022Continuous}. These architectures employ spatiotemporal structures across time frames rather than treating each spectrogram independently. As noted in \cite{GaoJSEN2021}, automotive mmWave FMCW radars operating at 77--81,GHz provide range, velocity, and angle estimates largely independent of weather and lighting, yet compact, low-cost arrays suffer from limited angular resolution. Because camera/lidar sensing degrades at night and in adverse weather, a key goal is to approach optical-level semantic perception using inexpensive radars.\blfootnote{\thanks{This research was supported by NSERC (Canada) under Grant Nos. RGPIN-2025-05010 and DGECR-2025-00215.}} 

Before deep learning became dominant, conventional multiclass target pipelines relied on handcrafted time–frequency descriptors and classical classifiers. Typical steps included time–frequency transforms, i.e., STFT \cite{Hanif2022}, extraction of physically meaningful features, e.g., Doppler centroid, rotating or vibrating structures, and then classification applied with shallow learners, e.g., KNN, SVM, because of their low-cost or low-power deployments and ability to generalize across sensors when features are stable. Besides, \cite{She2021SpecAug} focused on data-centric strategies, such as targeted spectrogram augmentations, to mitigate the challenges of small sample regimes in practical systems. However, challenges also arise when analog-to-digital converted (ADC) radar data are transformed into a 3D range–velocity–angle (RVA) heatmap cube per frame; the spatial separation in angle is often ambiguous on small arrays, so temporal motion cues across multiple frames are crucial.

With the advent of end-to-end learning, CNN-based approaches have become alternative solutions, where FMCW micro-Doppler representations improve activity classification on mmWave sensors. To better capture temporal dynamics of MDS, \cite{Zhu2022Continuous,Hsu2023Fall} applied time series models, e.g., LSTM, which proved effective for continuous recognition and fall detection. Related to multi-view learning, \cite{GaoJSEN2021} proposed RAMP-CNN, which avoids the computational burden of MDS data by slicing each 3D RVA cube into three orthogonal 2D sequences: range–angle (RA), range–velocity (RV), and velocity–angle (VA), then processing them with parallel lightweight CNN branches to detect the targets.

Despite strong progress with CNN-based methods, two limitations remain for MDS-based multiclass target classification. First, most CNN pipelines treat each spectrogram as an \emph{image} and do not explicitly model the \emph{time series} nature of MDS sequences across frames. The temporal coherence and inter-frame kinematic constraints are often ignored \cite{Abdu2022FF}. Second, the input representation in \cite{Zhu2022Continuous,Kim2022RTD} is the time–Doppler domain, whereas practical mmWave sensing yields richer information such as the angle and range domains, which can discard separability cues that are crucial when signatures overlap in multi-target scenes.
Therefore, with the MDS-based multiclass target classifier, we focus on (i) treating MDS as a \emph{sequence} by learning on stacked spatiotemporal tensors; (ii) preserving radar-native axes (range, Doppler, angle) in the input to maintain geometric correspondences; and (iii) injecting \emph{mobility-aware} constraints so that the network learns to separate and track class-specific motion patterns under target overlaps and viewpoint changes.
\begin{itemize}
    \item We introduce an end-to-end pipeline for MDS-based multiclass target classification of mobile targets using mmWave FMCW radar, tailored to multi-target scenes with overlapping signatures. The MDS data is analyzed with probabilistic structure to improve classification accuracy.
    \item We design an MDS-based transformer solution built on Vision Transformer (ViT) primitives, termed Temporal MDS-ViT (T-MDS-ViT), that ingests RVA sequences via patch embeddings and cross-axis attention, enabling multi-target disentanglement.
    \item We demonstrate data efficiency and deployable real-time execution when the model is trained with limited labeled data, fewer parameters and achieves accuracy comparable to heavier CNN baselines.
    \item We introduce an explainable mechanism where the attention layers of the T-MDS-ViT model can focus on the characteristics of the MDS data to enhance target classification.
\end{itemize}
Due to space limitations, the full-size high-resolution figure, NN architecture, data visualizations, AI/ML configurations, and training details are available in our \textit{code repository} at \url{github.com/TND-Lab/Temporal-MDS-ViT-Classification}
\section{System and Signal Modelling}
\subsection{System Components}
From the real measurement \cite{GaoJSEN2021}, the system topology considers a single radar transmitting a Frequency Modulated Continuous Wave (FMCW) waveform from $N_{\rm tx, ant} = 2$ antennas to multiple targets $K_{\mathrm{tar}} \geq 1$, then echoing back to the radar with $N_{\rm rx, ant} = 4$ antenna receivers. Let \(\theta \in \{\theta_{\mathrm{ZOD}}, \theta_{\mathrm{ZOA}}\}\) represent the zenith angles of departure and arrival, and \(\phi \in \{\phi_{\mathrm{AOD}}, \phi_{\mathrm{AOA}}\}\) represent the azimuth angles of departure and arrival, respectively. The transmit array response vector is \(\vec{\alpha}_{\mathrm{tx}} (\theta, \phi) = [\alpha_{\mathrm{tx}, 1} (\theta, \phi), \dots, \alpha_{\mathrm{tx}, N_{\mathrm{tx, ant}}} (\theta, \phi)]\), and the BS array response vector is \(\vec{\alpha}_{\mathrm{rx}} (\theta, \phi) = [\alpha_{\mathrm{rx}, 1} (\theta, \phi), \dots, \alpha_{\mathrm{rx}, N_{\mathrm{rx, ant}}} (\theta, \phi)]\).
\subsection{Signal modelling}
\paragraph{FMCW receiver signal}
Consider the transmitted complex sequence to be the linear FMCW chirp passband signal in the time domain $s_{\mathrm{tx}}(t)$, it can be presented as
\begin{align}
s_{\mathrm{tx}}(t)=\exp\!\left\{j2\pi\!\left(f_c t+\tfrac{S}{2}t^{2}\right)\right\},
\end{align}
where $t\in[0,T_{\rm c})$, \(T_{\rm c}\) is the chirp duration, \(f_c\) is the carrier frequency, \(S=\frac{B}{T_{\rm c}}\) is the slope, and \(B\) is the bandwidth. Assuming each \(i\)-th target has its physical measurements, with the range \(r_i(t)\), radial velocity \(v_i(t)\), and azimuth \(\theta_i(t)\) over slow time. At the receiver, the received signal $y(t,m, n_{\mathrm{tx}}, n_{\mathrm{rx}})$ after dechirping and low-pass filtering is the baseband beat at the antenna element pair \((n_{\mathrm{tx}}, n_{\mathrm{rx}})\), where $n_{\mathrm{tx}} \in \{1, \ldots, N_{\mathrm{tx, ant}}\}, n_{\mathrm{rx}} \in \{1, \ldots, N_{\mathrm{rx, ant}}\}$, and the slow-time chirp index \(m_{\rm c} \in \{0, 1, \ldots, M_{\rm c}-1\}\) can be presented as
\begin{align}
&y(t, m_{\rm c}, n_{\mathrm{tx}}, n_{\mathrm{rx}}) \!=\! \textstyle\sum_{i=1}^{K_{\mathrm{tar}}}\alpha_{\mathrm{tx}, n_{\mathrm{tx}}}\alpha_{\mathrm{rx}, n_{\mathrm{rx}}}
\exp\{j2\pi \\
&\times ((S\tau_i+f_{D,i})t \nonumber+ f_{D,i} m_{\rm c}T_{\rm r})\} \!+\!  w(t,m_{\rm c}, n_{\mathrm{tx}}, n_{\mathrm{rx}}), \nonumber
\end{align}
where $M_{\rm c}$ is the number of chirps per frame, \(\tau_i=2r_i/c\), \(f_{D,i}=2v_i/\lambda\), \(T_{\rm r}\) is the chirp repetition interval, and \(w(t,m, n_{\mathrm{tx}}, n_{\mathrm{rx}}) \in \mathcal{CN}(0, 1)\) is the noise factor.
\paragraph{Discrete sampling}
Each chirp sample can be discretized to $N_{\rm s}$ samples, where \((T_{\rm s} = 1/F_{\rm s})\) is the sampling period, with the ADC operating at rate \(F_{\rm s}\). The fast-time samples within each chirp are indexed by \(n_{\rm s}=\{0,\ldots,N_{\rm s}-1\}\), where \(N_{\rm s} =  T_{\rm c} \times F_{\rm s}\). Consider $t = n_{\rm s}T_{\rm s}$, where $y[n_{\rm s},m_{\rm c}, n_{\mathrm{tx}}, n_{\mathrm{rx}}]$ is the discrete received signal in the time domain of sample $N_{\rm s}$, it can be demonstrated as
$
    y[n_{\rm s},m_{\rm c}, n_{\mathrm{tx}}, n_{\mathrm{rx}}]
\!=\! \textstyle \sum_{i=1}^{K_{\mathrm{tar}}}\alpha_{\mathrm{tx}, n_{\mathrm{tx}}}\alpha_{\mathrm{rx}, n_{\mathrm{rx}}}
 \exp\{j2\pi((S\tau_i+f_{D,i})n_{\rm s}T_{\rm s} + f_{D,i} m_{\rm c}T_{\rm r})\} \!+\!  w[n_{\rm s},m_{\rm c}, n_{\mathrm{tx}}, n_{\mathrm{rx}}]
$.
In the TDM-MIMO system, each transmitter is activated in a dedicated time slot within the pulse repetition interval (PRI), which means that the slow-time sequence corresponding to the antenna transmitter \(n_{\mathrm{tx}}\) is shifted by an offset \(\Delta t_{n_\mathrm{tx}} = \frac{(n_{\rm tx} -1)T_{\rm r}}{N_{\rm tx, ant}}\). To form the virtual array, the received data must be reorganized into per-TX slow-time sequences and corrected for the slot-dependent delay. The response signal becomes
\begin{align} \label{eq_discrete_response}
    &y[n_{\rm s},m_{\rm c}, n_{\mathrm{tx}}, n_{\mathrm{rx}}]
    \!=\!\!\sum_{i=1}^{K_{\mathrm{tar}}}\alpha_{\mathrm{tx}, n_{\mathrm{tx}}}\alpha_{\mathrm{rx}, n_{\mathrm{rx}}} \!\!
\exp\{j2\pi \big((S\tau_i\!+\!f_{D,i}) \nonumber\\
&\times n_{\rm s}T_{\rm s} \!+\! f_{D,i} (m_{\rm c}T_{\rm r} - \Delta t_{n_\mathrm{tx}})\big)\} \!+\!  w[n_{\rm s},m_{\rm c}, n_{\mathrm{tx}}, n_{\mathrm{rx}}], 
\end{align}
\subsection{Range-Velocity-Angle Generation}
From the observation data in \eqref{eq_discrete_response}, with the main purpose of detecting and classifying the target, we applied the micro-Doppler spectrogram (MDS) process, which transforms the response signal into the range angle velocity time domain and can enhance the characteristics of multiple moving targets.
\paragraph{Range FFT process}
In the response signal presentation, the number of samples per chirp $N_{\rm s}$ represents the fast time domain. Let $Y[k_{\rm r}, m_{\rm c}, n_{\mathrm{tx}}, n_{\mathrm{rx}}]$ be the received signal in the frequency domain after applying the Fast Fourier Transform (FFT) in the fast time domain at range bin index $k_{\rm r} \in \{0, \ldots, N_{\rm s}-1\}$, antenna pair $(n_{\mathrm{tx}}, n_{\mathrm{rx}})$, and slow chirp index $m_{\rm c}$. Consider the Hamming window $\vec{w}_r \in \mathbb{R}^{1\times N_{\rm s}}$, with its values $\vec{w}_r[n] = 0.5(1 - \cos(2\pi \frac{n}{N_{\rm s}}))$, which reduces spectral leakage. The received signal in the fast time domain can be presented as
\begin{align}
    Y_{\rm r}[k_{\rm r}, m_{\rm c}, n_{\mathrm{tx}}, n_{\mathrm{rx}}] = &\sum_{n_{\rm s}=0}^{N_{\rm s}-1} y[n_{\rm s},m_{\rm c}, n_{\mathrm{tx}}, n_{\mathrm{rx}}] \vec{w}_r[n_{\rm s}] \nonumber \\ 
    &\times \exp\{-j 2\pi k_{\rm r} \tfrac{n_{\rm s}}{N_{\rm s}}\}.
\end{align}
\paragraph{Doppler FFT process}
Consider the slow time domain with $M_{\rm c}$ chirps per frame. Let $Y_{\rm rd}[k_{\rm r}, k_{\rm v}, n_{\mathrm{tx}}, n_{\mathrm{rx}}]$ be the received signal in the frequency domain after applying FFT in the slow time domain at slow chirp index $k_{\rm v} \in \{0, \ldots, M_{\rm c}-1\}$, range bin index $k_{\rm r}$, and antenna pair $(n_{\mathrm{tx}}, n_{\mathrm{rx}})$, it can be presented as
$
    Y_{\rm rd}[k_{\rm r}, k_{\rm v}, n_{\mathrm{tx}}, n_{\mathrm{rx}}] = \sum_{m_{\rm c}=0}^{M_{\rm c}-1} Y_{\rm r}[k_{\rm r},m_{\rm c}, n_{\mathrm{tx}}, n_{\mathrm{rx}}] \exp\{-j 2\pi k_{\rm v} m_{\rm c} / M_{\rm c}\}.
$
Because \eqref{eq_discrete_response} considers the virtual array phase-consistent across transmitters, applying Doppler compensation corrects the phase shift across chirps. The received signal after compensation in the slow-time domain can be presented as
\begin{align} \label{eq_range_doppler}
Y^{\rm comp}_{\rm rd}[k_{\rm r}, k_{\rm v}, n_{\mathrm{tx}}, n_{\mathrm{rx}}] &= Y_{\rm rd}[k_{\rm r}, k_{\rm v}, n_{\mathrm{tx}}, n_{\mathrm{rx}}] \nonumber \\  
&\quad\times \exp\{j 2\pi \tfrac{k_{\rm v}}{M_{\rm c} T_{\rm r}} \Delta t_{n_\mathrm{tx}}\}.
\end{align}
\paragraph{Range-Doppler Estimation}
To enhance the impact of the moving targets, according to \eqref{eq_range_doppler}, we apply CFAR to the range-Doppler map to present potential targets in range and Doppler bins. Let $P_{\rm rd}[k_{\rm r}, k_{\rm v}]$ be the average power across antenna elements, which is 
\begin{align}
P_{\rm rd}[k_{\rm r}, k_{\rm v}] = \sum_{n_{\rm tx}=1}^{N_{\rm tx, ant}}\sum_{n_{\rm rx}=1}^{N_{\rm rx, ant}} \big|Y^{\rm comp}_{\rm rd}[k_{\rm r}, k_{\rm v}, n_{\mathrm{tx}}, n_{\mathrm{rx}}]\big|^2.
\end{align}
For each cell at the range-Doppler pair \((k_{\rm r}, k_{\rm v})\), we estimate the average power $P_{\rm avg}[k_{\rm r}, k_{\rm v}]$ from surrounding training cells $N_{\rm cells}$,
$
P_{\rm avg}[k_{\rm r}, k_{\rm v}] = \frac{1}{N_{\rm cells}} \sum_{(i,j) \in \mathcal{T}} P_{\rm rd}[i, j],
$
where \( \mathcal{T} \) denotes the set of training cells and \( N_{\rm cells} \) is the number of training cells. Applying the CFAR algorithm with the threshold $\lambda_{\rm thresh}[k_{\rm r}, k_{\rm v}] = \alpha_{\rm thresh} P_{\rm avg}[k_{\rm r}, k_{\rm v}]$, where \( \alpha_{\rm thresh} \) the threshold factor is determined by the desired false alarm rate. The set of detected range-Doppler bins is identified as
\begin{align}
\mathcal{D}_{\rm rd} = \{(k_{\rm r}, k_{\rm v}) : P_{\rm rd}[k_{\rm r}, k_{\rm v}] > \lambda_{\rm thresh}[k_{\rm r}, k_{\rm v}]\}.
\end{align}
Let $\mathcal{D}^{\rm grouped}_{\rm rd} = f_{\rm groups} (\mathcal{D}_{\rm rd})$ be the grouping function that groups close range and velocity pairs from the potential range-Doppler bins.
\paragraph{Angle FFT}  
For each detected and grouped range-Doppler bin \((\hat{k}_{\rm r}^{(i)}, \hat{k}_{\rm v}^{(i)}) \in \mathcal{D}^{\rm grouped}_{\rm rd}\), where $i \in \{1, 2, \ldots, K_{\rm clustered}\}$, $K_{\rm clustered}$ is the number of grouped pairs. From \eqref{eq_range_doppler}, we extract the corresponding antenna-domain snapshot vector from the range-Doppler cube to obtain $\mathbf{Y}_{\rm rd}[\hat{k}_{\rm r}^{(i)}, \hat{k}_{\rm v}^{(i)}] = [Y_{\rm rd}[\hat{k}_{\rm r}^{(i)}, \hat{k}_{\rm v}^{(i)}, 1, 1], \ldots, Y_{\rm rd}[\hat{k}_{\rm r}^{(i)}, \hat{k}_{\rm v}^{(i)}, N_{\rm tx, ant}, N_{\rm rx, ant}]]$. The angle-time windowing $\vec{w}_{\rm \theta} \in \mathbb{R}^{N_{\rm tx, ant} \times N_{\rm rx, ant}}$, with its values $\vec{w}_{\rm \theta}[n_{\rm a}] = 0.5(1 - \cos(2\pi \tfrac{n_{\rm a}}{N_{\rm \theta}}))$, is applied across the antenna elements before the angle FFT process to reduce the RVA presentation at
\begin{align}
    Y_{\rm RVA}[\hat{k}_{\rm r}^{(i)}, \hat{k}_{\rm v}^{(i)}, k_{\rm \theta}^{(i)}] = &\sum_{n_{\rm a}=0}^{N_{\rm a}-1} \mathbf{Y}_{\rm rd}^{\text{comp}}[\hat{k}_{\rm r}^{(i)}, \hat{k}_{\rm v}^{(i)}][n_{\rm a}]\vec{w}_{\rm \theta}[n_{\rm a}] \nonumber \\&\times \exp\{-j 2\pi n_{\rm a}k_{\rm \theta}^{(i)}  / N_{\rm \theta}\},
\end{align}
where $k_{\rm \theta}^{(i)} \in \{0, \ldots, N_{\rm \theta}-1\}$, $n_{\rm a} \in \{1, \ldots, N_{\rm a}\}, N_{\rm a} = N_{\rm tx, ant} + N_{\rm rx, ant}$, and $N_{\rm \theta}$ is the angle FFT size. The estimated angle for the detection is given by the index of the maximum magnitude in the angle spectrum 
\begin{align}
\hat{k}_{\rm \theta}^{(i)} = \arg\max_{k_{\rm \theta}^{(i)}} |X_{\theta}[\hat{k}_{\rm r}^{(i)}, \hat{k}_{\rm v}^{(i)}, k_{\rm \theta}^{(i)}]|.
\end{align}
Let $\mathcal{D}^{\rm grouped}_{\rm RVA}$ be the group of the potential range, velocity, and angle of the target, with $(\hat{k}_{\rm r}^{(i)}, \hat{k}_{\rm v}^{(i)}, \hat{k}_{\rm \theta}^{(i)}) \in \mathcal{D}^{\rm grouped}_{\rm RVA}$.
\subsection{Micro-Doppler Spectrogram Generation}
\paragraph{Crop the RVA targets}
Assuming $K_{\rm tar}$ potential targets can be estimated from $\mathcal{D}^{\rm grouped}_{\rm RVA}$. Call $\mathcal{D}_{\text{RVA}}^{\text{cent}} = \{( \overline{k}_{\rm r}^{(i)}, \overline{k}_{\rm v}^{(i)}, \overline{k}_{\rm \theta}^{(i)}) \mid i = 1, \ldots, K_{\rm tar} \}$ the set of cluster centroids of the potential targets, $K_{\rm tar} < K_{\rm clusted}$. Applying the clustering function $f_{\rm clus}(\hat{k}_{\rm r}, \hat{k}_{\rm v}, \hat{k}_{\rm \theta})$ that predicts the cluster centroid indices $(\overline{k}_{\rm r}, \overline{k}_{\rm v}, \overline{k}_{\rm \theta})$, \((\overline{k}_{\rm r}, \overline{k}_{\rm v}, \overline{k}_{\rm \theta}) = f_{\rm clus}(\hat{k}_{\rm r}, \hat{k}_{\rm v}, \hat{k}_{\rm \theta})\). Because the cluster centroids have $Y_{\mathrm{RVA}}( \overline{k}_{\rm r}^{(i)}, \overline{k}_{\rm v}^{(i)}, \overline{i}_\theta^{(i)}) \in \mat{Y}_{\mathrm{RVA}}$, where $\mat{Y}_{\mathrm{RVA}} \in \mathbb{C}^{N_{\rm s} \times N_{\theta} \times M_{\rm c}}$, we crop the bounding boxes that contain the potential targets in the range-angle dimension. Call $\mat{Y}_{\mathrm{RVA}}^{(i),\rm bbox} \in \mathbb{C}^{N^{\rm res}_{\rm s} \times N^{\rm res}_{\rm \theta} \times M_{\rm c}}$ the cropped matrix containing the $i$-th potential target, with $\mat{Y}_{\mathrm{RVA}} \rightarrow \mat{Y}_{\mathrm{RVA}}^{(i),\rm bbox}$, where $N^{\rm res}_{\rm s} \leq N_{\rm s},  N^{\rm res}_{\rm \theta} \leq N_{\rm \theta}$ are the defined resolutions of the range and Doppler that contain the target, respectively. Because $M_{\rm c}$ chirps capture 1 frame of $\mat{Y}_{\mathrm{RVA}}^{(i),\rm bbox}$, for $K_{\rm frame}$, the cropped matrix containing the $i$-th potential target is 
\begin{align}
    \mat{Y}_{\mathrm{RVA, cube}}^{(i),\rm bbox} \!\!=\!\! [\mat{Y}_{\mathrm{RVA}, 1}^{(i),\rm bbox}, \mat{Y}_{\mathrm{RVA}, 2}^{(i),\rm bbox}, \ldots, \mat{Y}_{\mathrm{RVA}, K_{\rm frame}}^{(i),\rm bbox}]^{\sf T},
\end{align}
where $\mat{Y}_{\mathrm{RVA, cube}}^{(i),\rm bbox} \in \mathbb{C}^{N^{\rm res}_{s} \times N^{\rm res}_{\theta} \times M_{\rm c} \times K_{\rm frame}}$. Considering the time dimension $N_{\rm time} = M_{\rm c} \times K_{\rm frame}$, the data cube is reshaped to $
    \mat{Y}_{\mathrm{RVA, cube}}^{(i),\rm bbox}: \mathbb{C}^{N^{\rm res}_{s} \times N^{\rm res}_{\theta} \times M_{\rm c} \times K_{\rm frame}} \mapsto \mathbb{C}^{N^{\rm res}_{s} \times N^{\rm res}_{\theta} \times N_{\rm time}}.
$
\paragraph{Short-time Fourier transform process}
Consider the $i$-th potential target has $\vec{y}_{\mathrm{RVA, cube}}^{(i)}[n_{\rm s}, n_{\theta}] \in \mathbb{C}^{1 \times N_{\rm time}}$ with 
\(n_{\rm s}\in\{0,\ldots,N^{\rm res}_{s}-1\}\), 
\(n_{\rm \theta}\in\{0,\ldots,N^{\rm res}_{\theta}-1\}\), $\vec{y}_{\mathrm{RVA, cube}}^{(i)}[n_{\rm s}, n_{\theta}] \in \mat{Y}_{\mathrm{RVA, cube}}^{(i),\rm bbox}$. Apply a short-time Fourier transform (STFT) along the slow-time axis for each spatial cell \((n_{\rm s},n_{\rm \theta})\).  
Let \(\vec{w}_{\rm v} \in \mathbb{C}^{1\times M_{\rm c}}\) denote the analysis Hamming window, with \(H_{\rm hop} = M_{\rm c} - N_{\rm overlap}\) the hop size, $N_{\rm overlap}$ the number of sample overlaps, and \(N_{\rm fft} = M_{\rm c}\) the STFT FFT length. For the signal \(\vec{y}_{\mathrm{RVA, cube}}^{(i)}[n_{\rm s}, n_{\theta}]\), for each time step $t = m_{\rm fft} H_{\rm hop} + m_{\rm c}$, the STFT at frequency bin \(m_{\rm c}\) and frame index \(m_{fft}\) is
\begin{align}
&S_i[n_{\rm s},n_{\rm \theta}, \,m_{\rm c},m_{\rm fft}]
\!=\! \!\sum_{m_{\rm c}=0}^{N_{\rm fft}-1}\! \vec{y}_{\mathrm{RVA, cube}}^{(i)}[n_{\rm s}, n_{\theta}]  \nonumber \\ 
&\times [m_{\rm fft} H_{\rm hop} \!+\! m_{\rm c}] \vec{w}_{\rm v}[m_{\rm c}]\;
\exp\{-j2\pi m_{\rm c} \frac{m_{\rm fft}}{N_{\rm fft}}\},
\label{eq:stft_cell}
\end{align}
where the frame indices are \(m_{\rm fft}=\{0,\ldots,N_{\rm fft}-1\}\) such that \(m_{\rm fft} H_{\rm hop} + m_{\rm c}\le N_{\rm time}\), and the frequency bins are \(m_{\rm c}\in\{0,\ldots,N_{\rm fft}-1\}\). The resulting spectrogram for the whole cropped cube is the micro-Doppler signature (MDS) data of the $i$-th potential target, $\mathbf{S}_i \in \mathbb{C}^{\,N^{\rm res}_{\rm s}\times N^{\rm res}_{\rm \theta}\times N_{\rm fft}\times N_{\rm fft}}.$
\subsection{Problem Formulation}
From the MDS data of the $i$-th potential target, $\mathbf{S}_i \in \mathbb{C}^{N^{\mathrm{res}}_{\rm s} \times N^{\mathrm{res}}_{\rm \theta} \times N_{\mathrm{fft}} \times N_{\mathrm{fft}}}$, let $c_{i} \in \{0, 1, \ldots, K_{\mathrm{tar}}\}$ be the target class to classify from the observation of the MDS data. Consider the posterior probability $p(y_i|\mathbf{S}_i)$ corresponding to the $i$-th class. For the maximum posterior estimation problem, the formulation can be expressed as:
\begin{subequations}
\begin{alignat}{2}
\hat{c}_i &= \arg \max_{k \in \{0,1,\ldots,K_{\mathrm{tar}}\}} p(c_i = k \mid \mathbf{S}_i) \label{eq_problem_formulation} \\
&\text{subject to} \quad 
    N^{\mathrm{res}}_{s} \le N_{\rm s}, 
    \quad N^{\mathrm{res}}_{\theta} \le N_{\theta} \\
&\hspace{1.7cm} \Delta r = \frac{c F_{\rm s}}{2 S N_{\rm s}}, 
    \quad r_{\max} = \frac{c F_{\rm s}}{2 S} \\
&\quad \Delta v = \frac{\lambda}{2 T_{\rm c} N_c}, 
    \quad v_{\max} = \frac{\lambda}{4 T_{\rm c}}, 
    \quad \Delta \theta = \frac{2}{N_{\theta}} .
\end{alignat}
\end{subequations}
The problem in \eqref{eq_problem_formulation} is not easy to solve because the MDS has multi-target overlap from bins, and the effect of the temporal coherence factor should be considered. Applying traditional machine learning seems impossible with the geometry-rich inputs. Furthermore, the traditional CNN model is limited to three dimensions, while the MDS data are four-dimensional. Our proposed model uses the attention mechanism to handle the temporal information while reducing the dimensionality of the MDS data.
\section{Proposed Pipeline for ViT-based Solution}

\subsection{Analysis of Probabilistic Features}
\begin{figure}[t]
    \centering
    \includegraphics[width=.95\linewidth]{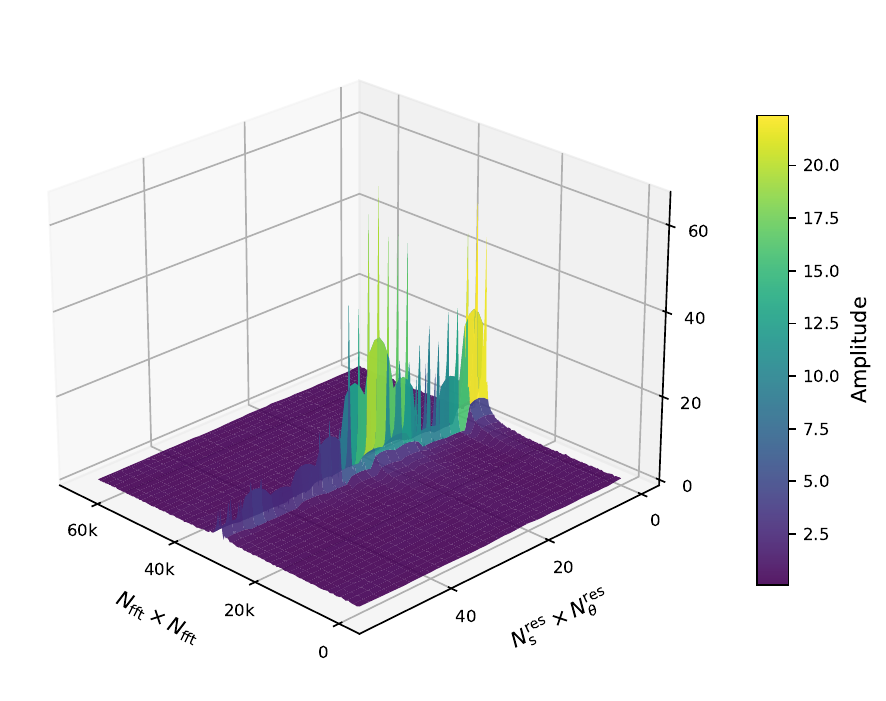}
        \label{fig_mds_car}
    \caption{
    The distribution of the MDS of the car class. Depending on different locations, speeds, and directions, the distribution of each target is different.}
    \label{fig_mds_all}
\end{figure}
According to \eqref{eq_problem_formulation}, applying the Bayes formulation, $p(c_i = k|\mathbf{S}_i) =\frac{p(\mathbf{S}_i|c_i= k) p(c_i=k)}{p(\mathbf{S}_i)}$. Consider the likelihood probability $p(\mathbf{S}_i|c_i= k) = \sum_{n_{\rm s}, m_{\rm c}, n_{\rm \theta}}p(\mathbf{S}_i|r_{n_{\rm s}}, v_{m_{\rm c}}, \theta_{n_{\rm \theta}}, c_i=k) p(r_{n_{\rm s}}, v_{m_{\rm c}}, \theta_{n_{\rm \theta}}|c_i= k)$, where $n_{\rm s} \in \{0, 1, \ldots, N_{\rm s}^{\rm res}-1\}, m_{\rm c} \in \{0, 1, \ldots, M_{\rm c}-1\}, n_{\rm \theta} \in \{0, 1, \ldots, N_{\theta}^{\rm res}-1\}$. Because the physics of measurement given $(r_{n_{\rm s}}, v_{m_{\rm c}}, \theta_{n_{\rm \theta}})$ is independent of the class label $c_i = k$, $p(\mathbf{S}_i|r_{n_{\rm s}}, v_{m_{\rm c}}, \theta_{n_{\rm \theta}}, c_i=k) = p(\mathbf{S}_i|r_{n_{\rm s}}, v_{m_{\rm c}}, \theta_{n_{\rm \theta}})$. Assuming the prior distribution of $(r_{n_{\rm s}}, v_{m_{\rm c}}, \theta_{n_{\rm \theta}})$ is a uniform distribution following the constraint resolutions, so $p(\mathbf{S}_i|c_i= k) = \frac{1}{G_k} \sum_{(n_{\rm s}, m_{\rm c}, n_{\rm \theta}) \in G_k}p(\mathbf{S}_i|r_{n_{\rm s}}, v_{m_{\rm c}}, \theta_{n_{\rm \theta}})$, where $G_k$ is the set of resolutions. The problem formulation becomes
\begin{align} \label{eq_problem_formulation_2}
    p(c_i = k|\mathbf{S}_i) \propto p(c_i = k)\frac{1}{G_k} \sum_{(n_{\rm s}, m_{\rm c}, n_{\rm \theta}) \in G_k}p(\mathbf{S}_i|r_{n_{\rm s}}, v_{m_{\rm c}}, \theta_{n_{\rm \theta}}) 
\end{align}
As can be seen in Fig.~\ref{fig_mds_all}, the distribution of the MDS data of the target car is focused on the center line, which is affected by the changes of the physical parameters for $r_{n_{\rm s}}, v_{m_{\rm c}}, \theta_{n_{\rm \theta}}$ of mobility targets over time.

\subsection{Proposed T-MDS-ViT Classifier}
Because the MDS data has 4 dimensions, $\mathbf{S}_i \in \mathbb{C}^{N^{\rm res}_{\rm s}\times N^{\rm res}_{\rm \theta}\times N_{\rm fft}\times N_{\rm fft}}$, applying traditional ML, i.e., MAP or MLE, is difficult because of the large number of features. The CNN model is a potential solution; however, MDS has more than 3 dimensions, and the temporal characteristics of MDS have long-range dependencies, while the conventional CNNs are naturally local feature extractors.
\subsubsection{Input Dimension}
In the MDS data $\mathbf{S}_i$, the dimensions $N^{\rm res}_{\rm s}\times N^{\rm res}_{\rm \theta}$ represent the bins of range-angle, each bin being presented by the STFT heatmap $N_{\rm fft}\times N_{\rm fft}$. Call $f_{\rm [dim]}(.): \mathbb{C}^{N^{\rm res}_{s}\times N^{\rm res}_{\theta}\times N_{\rm fft}\times N_{\rm fft}} \mapsto \mathbb{R}^{N^{\rm res}_{\rm bin} \times N_{\rm fft}\times N_{\rm fft}}$ the reduced-dimension function, where $N^{\rm res}_{\rm bin} = N^{\rm res}_{s}\times N^{\rm res}_{\theta}$ represents the number of slice bins. Because the MDS data are complex and not suitable for the proposed model, we applied the absolute values of the MDS data to obtain $\mathbf{S}^{\rm dim}_i = f_{\rm [dim]}(|\mathbf{S}_i|) \in \mathbb{R}^{N^{\rm res}_{\rm bin} \times N_{\rm fft}\times N_{\rm fft}}$. Applying this to \eqref{eq_problem_formulation_2} and \eqref{eq_problem_formulation}, the problem formulation becomes
\begin{align}
    \hat{c}_i = \arg \max_{k \in \{0, 1, \ldots, K_{\mathrm{tar}}\}} p(c_i = k|\mathbf{S}^{\dim}_i).
\end{align}
\subsubsection{Model Architecture}
Based on the works of \cite{Dosovitskiy2020VIT}, the Vision Transformer (ViT) provides a powerful architecture for classification tasks on spatio-temporal inputs. In this work, we propose a T-MDS-ViT to model the classification function \(f_{\mathrm{vit}}\), parameterized by \(\boldsymbol{\Phi}\). The input of \(f_{\mathrm{vit}}\) is the dimension-reduced MDS data \(\mathbf{S}^{\dim}_i \in \mathbb{R}^{N^{\mathrm{res}}_{\mathrm{bin}} \times N_{\mathrm{fft}} \times N_{\mathrm{fft}}}\), where \(N^{\mathrm{res}}_{\mathrm{bin}}\) denotes the number of temporal slices, and each slice corresponds to a two-dimensional STFT heatmap. As can be seen in Fig.~\ref{fig_mds_all}, the order of bins is arranged to follow the split patch characteristics of the ViT model. Because the bins of range-angles $N_{\rm bin}^{\rm res}$ are ordered in the temporal domain, they can be considered as patches in the T-MDS-ViT model. Call $f_{[\rm patch]}(.): \mathbb{R}^{N^{\mathrm{res}}_{\mathrm{bin}} \times N_{\mathrm{fft}} \times N_{\mathrm{fft}}} \mapsto \mathbb{R}^{N^{\mathrm{res}}_{\mathrm{bin}} \times N_{\mathrm{emb}}}$ the patch embedding function, which flattens each STFT heatmap into a vector of length $N_{\mathrm{fft}} \times N_{\mathrm{fft}}$ and projects it into the embedding dimension $N_{\mathrm{emb}}$ via the updated parameter $\boldsymbol{\Phi}_{\rm [patch]}$. It is noted that the positional embedding $\vec{P}_{\mathrm{cls}} \in \mathbb{R}^{N^{\mathrm{res}}_{\mathrm{bin}} \times N_{\mathrm{emb}}}$ is appended to the embedding sequence. The transformer encoder network of ViT is constructed from $N_{\rm blocks}$ stacked transformer blocks. Each block $f_{\mathrm{vit}}^{[\mathrm{bl}_{\ell}]}$ consists of
$
f_{\mathrm{vit}}^{[\mathrm{bl}_{\ell}]} = f_{\mathrm{vit}}^{[\mathrm{res2}_{\ell}]} \circ f_{\mathrm{vit}}^{[\mathrm{mlp}_{\ell}]} \circ f_{\mathrm{vit}}^{[\mathrm{res1}_{\ell}]} \circ f_{\mathrm{vit}}^{[\mathrm{mha}_{\ell}]},
$
updated by the parameter $\boldsymbol{\Phi}_{[\mathrm{bl}_{\ell}]}$ where $\ell \in \{1, \ldots, N_{\rm blocks}\}$, $f_{\mathrm{vit}}^{[\mathrm{mha}_{\ell}]}$ denotes the multi-head self-attention layer, $f_{\mathrm{vit}}^{[\mathrm{mlp}_{\ell}]}$ the feed-forward MLP with expansion ratio $r_{\mathrm{mlp}}$, and $f_{\mathrm{vit}}^{[\mathrm{res1}_{\ell}]}, f_{\mathrm{vit}}^{[\mathrm{res2}_{\ell}]}$ are residual connections with layer normalization. After $N_{\mathrm{blocks}}$ transformer blocks, the updated features are passed to the classification head $f_{\mathrm{vit}}^{[\mathrm{head}]}$, which is a linear projection followed by a softmax activation. The final classification function is therefore given by
\begin{align} \label{eq_math_vit_comp}
    &f_{\mathrm{vit}}(\mathbf{S}^{\dim}_i; \boldsymbol{\Phi}) 
    \!=\! f_{\mathrm{vit}}^{[\mathrm{head}]}\Big(
        f_{\mathrm{vit}}^{[\mathrm{bl}_L]}\big(
            f_{\mathrm{vit}}^{[\mathrm{bl}_{L-1}]}\big(
                \cdots 
                f_{\mathrm{vit}}^{[\mathrm{bl}_1]}\big(
                    f_{[\mathrm{patch}]}\big( \nonumber \\ &f_{[\mathrm{dim}]}(\mathbf{S}^{\dim}_i); \boldsymbol{\Phi}_{[\mathrm{patch}]}\big) 
                    \!+\! \vec{P}_{\mathrm{cls}}
                    ; \boldsymbol{\Phi}_{[\mathrm{bl}_1]}
                \big)
            ; \boldsymbol{\Phi}_{[\mathrm{bl}_{L-1}]}
            \big)
        ; \boldsymbol{\Phi}_{[\mathrm{bl}_L]}
    \Big)
    .
\end{align}
Denote $\hat{c}_i = f_{\mathrm{vit}}(\mathbf{S}^{\dim}_i; \boldsymbol{\Phi})$ the estimated target class. The problem formulation becomes a multi-class classification task: $\hat{c}_i = \arg \max_{k}p(c_i = k \mid \mathbf{S}^{\dim}_i; \boldsymbol{\Phi})$.
\subsubsection{Loss Function}
To train the parameters $\boldsymbol{\Phi}$ of the proposed T-MDS-ViT model, we employ the cross-entropy loss, which measures the divergence between the predicted distribution and the ground-truth labels. 
For the $i$-th training sample $(\mathbf{S}^{\dim}_i, c_i)$, the predicted probability of class $k$ is
\begin{align}
    p(c_i = k \mid \mathbf{S}^{\dim}_i; \boldsymbol{\Phi})
    = \frac{\exp\!\left(u_k(\mathbf{S}^{\dim}_i; \boldsymbol{\Phi})\right)}{\sum_{j=1}^{K_{\mathrm{tar}}} \exp\!\left(u_j(\mathbf{S}^{\dim}_i; \boldsymbol{\Phi})\right)},
\end{align}
where $u_k(\mathbf{S}^{\dim}_i; \boldsymbol{\Phi})$ denotes the logit for class $k$, and $K_{\mathrm{tar}}$ is the total number of target classes. The cross-entropy loss over the training dataset $\mathcal{D} = \{(\mathbf{S}^{\dim}_i, c_i)\}_{i=1}^{N_{\rm samples}}$, where $N_{\rm samples} = |\mathcal{B}|$, is then defined as
\begin{align}
    \mathcal{L}_{CE}(\boldsymbol{\Phi})
    \!\!=\!\! -\frac{1}{N} \sum_{i=1}^{N} \!\! \sum_{k=1}^{K_{\mathrm{tar}}} 
    \mathbf{I}(c_i \!\!=\!\! k)\log p(c_i \!=\! k \!\!\mid\!\! \mathbf{S}^{\dim}_i; \boldsymbol{\Phi}),
\end{align}
where $\mathbf{I}(c_i = k)$ is the indicator function that equals $1$ if the ground-truth class of the $i$-th sample is $k$, and $0$ otherwise. 
From the distribution of the MDS data in Fig.~\ref{fig_mds_all}, we observed that the values of the MDS focus on specific regions. To enhance feature selection in the proposed model, the L2 regularization term is incorporated into the loss function $\mathcal{L}_{CE}(\boldsymbol{\Phi})$ to penalize large values of $ \boldsymbol{\Phi}$, thereby encouraging sparsity or smaller weights for less important features. The total loss can be described as
\begin{align}
    \mathcal{L}(\boldsymbol{\Phi}) = \mathcal{L}_{CE}(\boldsymbol{\Phi}) + \lambda \|\boldsymbol{\Phi}\|^2_2,
\end{align}
where $\lambda = 0.01$ is the weight decay. Because the optimal problem $\hat{c}_i = \arg \max_{k}p(c_i = k \mid \mathbf{S}^{\dim}_i; \boldsymbol{\Phi})$ is the reverse of the cross-entropy loss, the optimal parameters are obtained by minimizing the loss: $\boldsymbol{\Phi}^\star = \arg \min_{\boldsymbol{\Phi}} \; \mathcal{L}(\boldsymbol{\Phi})$.

\subsubsection{Backpropagation and Optimization}
To optimize the parameters $\boldsymbol{\Phi}$ of the proposed T-MDS-ViT, the gradients of the loss function $\mathcal{L}(\boldsymbol{\Phi})$ are computed with respect to the model parameters using the backpropagation process
\begin{align}
\nabla_{\boldsymbol{\Phi}} \mathcal{L}(\boldsymbol{\Phi})
&= \nabla_{\boldsymbol{\Phi}} \mathcal{L}_{CE}(\boldsymbol{\Phi})
\!+\! \lambda \frac{\partial}{\partial \boldsymbol{\Phi}} \|\boldsymbol{\Phi}\|_2^2 \nonumber \\
&= \nabla_{\boldsymbol{\Phi}} \mathcal{L}_{CE}(\boldsymbol{\Phi}) \!+\! 2 \lambda \boldsymbol{\Phi}.
\end{align}
To update the parameters efficiently, we employ the Adam optimizer, which combines adaptive learning rates with momentum-based updates. At the $t$-th iteration, for each parameter $\phi$, the gradient $g_t = \nabla_{\boldsymbol{\Phi}} \mathcal{L}(\boldsymbol{\Phi}_t)$ is used to compute the biased first and second moment estimates $m_t = \beta_1 m_{t-1} + (1 - \beta_1) g_t, v_t = \beta_2 v_{t-1} + (1 - \beta_2) g_t^2$, where $\beta_1, \beta_2 \in (0,1)$ are exponential decay rates for the moments. To correct for initialization bias, the unbiased estimates are $\hat{m}_t = \frac{m_t}{1 - \beta_1^t}, \hat{v}_t = \frac{v_t}{1 - \beta_2^t}$. Finally, the parameter update rule is given by
$
\boldsymbol{\Phi}_{t+1} = \boldsymbol{\Phi}_t - \eta \frac{\hat{m}_t}{\sqrt{\hat{v}_t} + \epsilon},
$
where $\eta$ is the learning rate and $\epsilon$ is a small constant ensuring numerical stability. The optimization process is iterated until convergence, yielding the optimal parameters $\boldsymbol{\Phi}^\star$ that minimize the total loss.
\subsubsection{Training Algorithm}
In our problem, the training set $\mathcal{D}_{\rm train}$ contains fewer than 100 MDS samples, which is very small. To ensure a fair comparison between our proposed model, T-MDS-ViT, and the CNN-based baseline, we employed cross-validation on $\mathcal{D}_{\rm train}$. The overall training pipeline is summarized in Algorithm~\ref{algo_temporal_vit_training}.
\begin{algorithm}[htbp]
\SetAlgoLined
\KwIn{MDS training dataset $\mathcal{D}_{\rm train} = \{(\mathbf{S}_i, c_i)\}_{i=1}^N$ where $\mathbf{S}_i \in \mathbb{C}^{N^{\rm res}_{\rm s}\times N^{\rm res}_{\rm \theta}\times N_{\rm fft}\times N_{\rm fft}}$ and $c_i \in \{0, 1, \ldots, K_{\mathrm{tar}}\}$, number of folds $K_{\rm fold}$}
\KwOut{Best model parameters $\boldsymbol{\Phi}^\star$, average accuracy $\mathrm{Acc}_{\rm avg}$}

\textbf{Initialization:} learning rate $\eta$, batch size $b$, regularization parameter $\lambda$, Adam optimizer parameters $\beta_1$, $\beta_2$, $\epsilon$, and number of epochs $K_{\mathrm{epoch}}$\;

\textbf{Cross-validation setup:} Randomly partition dataset $\mathcal{D}_{\rm train}$ into $K_{\rm fold}$ equal-sized folds $\{\mathcal{F}_1, \mathcal{F}_2, \ldots, \mathcal{F}_{K_{\rm fold}}\}$\;

Initialize performance metrics storage: $\{\mathrm{Acc}_j\}_{j=1}^{K_{\rm fold}}$\;

Initialize best accuracy $\mathrm{Acc}_{\text{best}} = 0$ and best parameters $\boldsymbol{\Phi}^\star$\;

\For{$j = 1, \dots, K_{\rm fold}$}{
    \textbf{Fold $j$ setup:}\;
    $\mathcal{D}_{\text{train}}^{(j)} \leftarrow \bigcup_{i \neq j} \mathcal{F}_i$;
    $\mathcal{D}_{\text{val}}^{(j)} \leftarrow \mathcal{F}_j$;
    
    \textbf{Initialize model:} Initialize parameters $\boldsymbol{\Phi}^{(j)}$, moment estimates $m_0 = 0$, $v_0 = 0$\;
    
    \For{$e = 1, \dots, K_{\mathrm{epoch}}$}{
        Shuffle training dataset $\mathcal{D}_{\text{train}}^{(j)}$\;
        
        \For{each batch $\mathcal{B}^{[e]} = \{(\mathbf{S}_i^{[e]}, c_i^{[e]})\}_{i=1}^{b}$ in $\mathcal{D}_{\text{train}}^{(j)}$}{
            
            \For{$i = 1, \dots, b$}{
                $\mathbf{S}^{\mathrm{dim}, [e]}_{i} \leftarrow f_{\rm dim}(|\mathbf{S}_i^{[e]}|) \in \mathbb{R}^{N^{\rm res}_{\rm bin} \times N_{\rm fft}\times N_{\rm fft}}$\;
            }
            
            $\{\mathbf{E}_{i}^{[e]}\}_{i=1}^{b} \leftarrow f_{\rm patch}(\{\mathbf{S}^{\mathrm{dim},[e]}_{i}\}_{i=1}^{b}) \in \mathbb{R}^{N^{\rm res}_{\rm bin} \times N_{\rm emb}}$\;
            
            $\{\mathbf{X}_i^{[0],[e]}\}_{i=1}^{b} \leftarrow \{\mathbf{E}_i^{[e]} + \vec{P}_{\mathrm{cls}}\}_{i=1}^{b}$\;
            
            \For{$\ell = 1, \dots, L$}{
                $\{\mathbf{X}_i^{[\ell],[e]}\}_{i=1}^{b} \leftarrow f_{\mathrm{vit}}^{[\mathrm{bl}_{\ell}]}(\{\mathbf{X}_i^{[\ell-1],[e]}\}_{i=1}^{b}; \boldsymbol{\Phi}_{[\mathrm{bl}_{\ell}]}^{(j)})$\;
            }
            
            $\{\hat{\mathbf{p}}_i^{[e]}\}_{i=1}^{b} \leftarrow f_{\mathrm{vit}}^{[\mathrm{head}]}(\{\mathbf{X}_i^{[L],[e]}\}_{i=1}^{b}; \boldsymbol{\Phi}_{[\mathrm{head}]}^{(j)})$\;
            
            $\mathcal{L}_{CE}^{[e]} \leftarrow -\frac{1}{b} \sum_{i=1}^{b} \sum_{k=1}^{K_{\mathrm{tar}}} \mathbf{I}(c_i^{[e]} = k)\, \log p(c_i^{[e]} = k \mid \mathbf{S}^{\mathrm{dim}, [e]}_{i}; \boldsymbol{\Phi}^{(j)})$\;
            
            $\mathcal{L}_{L2}^{[e]} \leftarrow \lambda \|\boldsymbol{\Phi}^{(j)}\|^2_2$\;
            
            $\mathcal{L}^{[e]} \leftarrow \mathcal{L}_{CE}^{[e]} + \mathcal{L}_{L2}^{[e]}$\;
            
            $g_t \leftarrow \nabla_{\boldsymbol{\Phi}^{(j)}} \mathcal{L}^{[e]} = \nabla_{\boldsymbol{\Phi}^{(j)}} \mathcal{L}_{CE}^{[e]} + 2\lambda \boldsymbol{\Phi}^{(j)}$\;
            
            $m_t \leftarrow \beta_1 m_{t-1} + (1 - \beta_1) g_t$\;
            
            $v_t \leftarrow \beta_2 v_{t-1} + (1 - \beta_2) g_t^2$\;
            
            $\hat{m}_t \leftarrow \frac{m_t}{1 - \beta_1^t}$, $\hat{v}_t \leftarrow \frac{v_t}{1 - \beta_2^t}$\;
            
            $\boldsymbol{\Phi}_{t+1}^{(j)} \leftarrow \boldsymbol{\Phi}_t^{(j)} - \eta \frac{\hat{m}_t}{\sqrt{\hat{v}_t} + \epsilon}$\;
        }
    }
    
    Evaluate $\mathcal{D}_{\text{val}}^{(j)}$ and compute $\mathrm{Acc}_j$\;
    
    \If{$\mathrm{Acc}_j > \mathrm{Acc}_{\text{best}}$}{
        $\mathrm{Acc}_{\text{best}} \leftarrow \mathrm{Acc}_j$;
        $\boldsymbol{\Phi}^\star \leftarrow \boldsymbol{\Phi}^{(j)}$\;
    }
}
$\mathrm{Acc}_{\text{avg}} \leftarrow \frac{1}{K_{\rm fold}} \sum_{j=1}^{K_{\rm fold}} \mathrm{Acc}_j$\;
\Return{$\boldsymbol{\Phi}^\star, \mathrm{Acc}_{\text{avg}}$}

\caption{Training Procedure of Temporal ViT for MDS-based multiclass target classification}
\label{algo_temporal_vit_training}
\end{algorithm}

\section{Evaluation Metrics}
\subsubsection{Accuracy}
The classification accuracy is defined as the ratio between the number of correctly classified samples and the total number of samples. Let $\mathbf{I}(\cdot)$ denote the indicator function. Then, the accuracy can be expressed as
$
    \mathrm{Acc} 
    = \frac{1}{\mathcal{D}_{\mathrm{test}}} 
    \sum_{i=1}^{\mathcal{D}_{\mathrm{test}}} 
    \mathbf{I}(\hat{c}_i = c_i),
$
where $\mathcal{D}_{\mathrm{test}}$ is the number of test samples, $c_i$ is the ground-truth class label of the $i$-th sample, and $\hat{c}_i$ is the predicted label obtained from \eqref{eq_problem_formulation}.
\subsubsection{Confusion Matrix}
We employed the confusion matrix $\mathbf{M} \in \mathbb{N}^{K_{\mathrm{tar}} \times K_{\mathrm{tar}}}$, which provides a comprehensive summary of classification outcomes across all classes, to measure the performance of each target based on our datasets. Let $K_{\mathrm{tar}}$ denote the total number of target classes. The value of each cell in $\mathbf{M}$ can be formed as 
$
    M_{p,q} = \sum_{i=1}^{\mathcal{D}_{\rm test}} \mathbf{I}(c_i = p) \mathbf{I}(\hat{c}_i = q),
$
where $M_{p,q}$ counts the number of samples whose true class is $p$ and predicted class is $q$. For a specific class $k \in \{1,\ldots,K_{\mathrm{tar}}\}$, the standard metrics can be derived from $\mathbf{M}$ as follows $\mathrm{TP}_k = M_{k,k}$ are correctly predicted samples of class $k$, $\mathrm{FP}_k = \sum_{\substack{p=1 \\ p \neq k}}^{K_{\mathrm{tar}}} M_{p,k}$ are incorrectly predicted as class $k$, $\mathrm{FN}_k = \sum_{\substack{q=1 \\ q \neq k}}^{K_{\mathrm{tar}}} M_{k,q}$ are class-$k$ samples, misclassified as another class, and $\mathrm{TN}_k = \sum_{\substack{p=1 \\ p \neq k}}^{K_{\mathrm{tar}}}  \sum_{\substack{q=1 \\ q \neq k}}^{K_{\mathrm{tar}}} M_{p,q}$ are neither belonging to class $k$ nor predicted as $k$.
\subsection{Explaination AI}
To demonstrate how the proposed T-MDS-ViT model can highlight the most discriminative regions of the MDS input, we employ a gradient-based explainable AI method, Grad-CAM, to analyze the interpretability of transformer embeddings based on our proposed AM4C \cite{Thinh_ICC_2025}. The method uses the backpropagated gradients to generate attention heatmaps, emphasizing the most salient (high-value) temporal–spectral bins in the representation $\mathbf{S}^{\dim}_i$. Denote the activation map at the $t$-th token and $j$-th embedding dimension from the $\ell$-th transformer block as
\begin{align}
    f_{\mathrm{vit}}^{[\mathrm{bl}_\ell]}(\mathbf{S}^{\dim}_i; \boldsymbol{\Phi}_{[\mathrm{bl}_\ell]})
= &\{ A_{t,j}^{(\ell)} \mid t \in \{1,\ldots,N^{\mathrm{res}}_{\mathrm{bin}}\}, \nonumber \\& j \in \{1,\ldots,N_{\mathrm{emb}}\}\}.
\end{align}
According to \cite[Eq. (20-21)]{Thinh_ICC_2025}, the activation map relevance score for a token $t$ is
\begin{align} \label{eq_explaination_ai}
    M_t^{(k)} = \mathtt{ReLU}\left(\frac{1}{N^{\mathrm{res}}_{\mathrm{bin}}}\sum_{t=1}^{N^{\mathrm{res}}_{\mathrm{bin}}}\sum_{j=1}^{N_{\mathrm{emb}}} \frac{\partial z_{i,k}}{\partial A_{t,j}^{(\ell)}} A_{t,j}^{(\ell)}\right),
\end{align}
where the gradient of the class score $z_{i,k}$ is taken with respect to each activation $\frac{\partial z_{i,k}}{\partial A_{t,j}^{(\ell)}}$. Mapping these values back to the original STFT bins yields a spatio-temporal visualization $\mathbf{M}_i^{(k)}$ that exposes the attention regions where the T-MDS-ViT focuses when performing target classification.
\section{Numerical Results and Explanation}
\begin{figure}[t]
	\centering
	\subfloat[]{%
		\includegraphics[width=.55\linewidth]{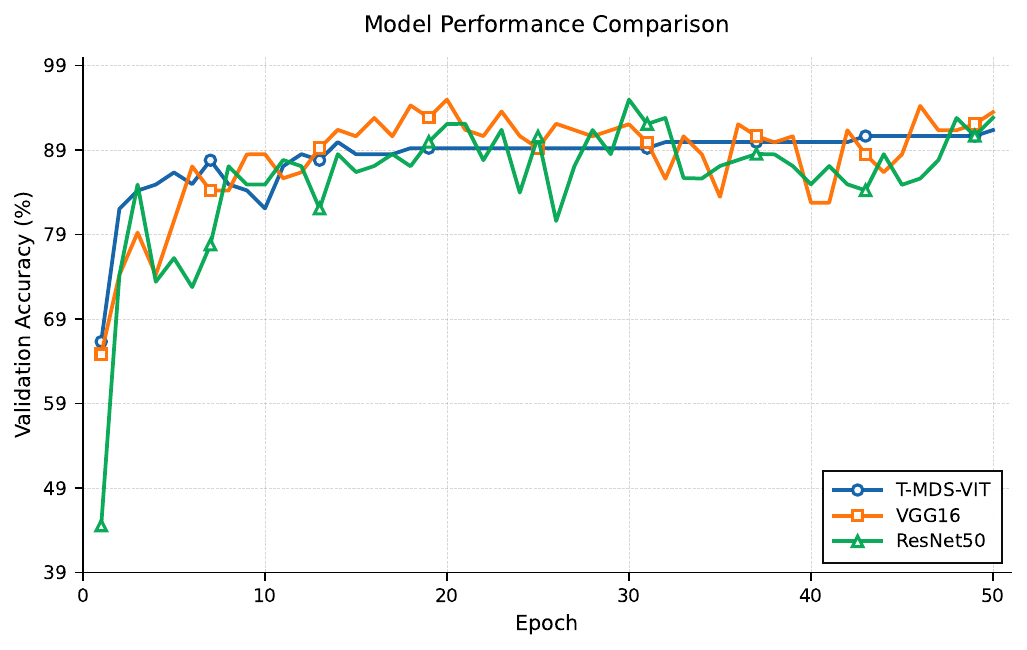}
		\label{fig_model_comparison}} \hfill
	\subfloat[]{%
		\includegraphics[width=.4\linewidth]{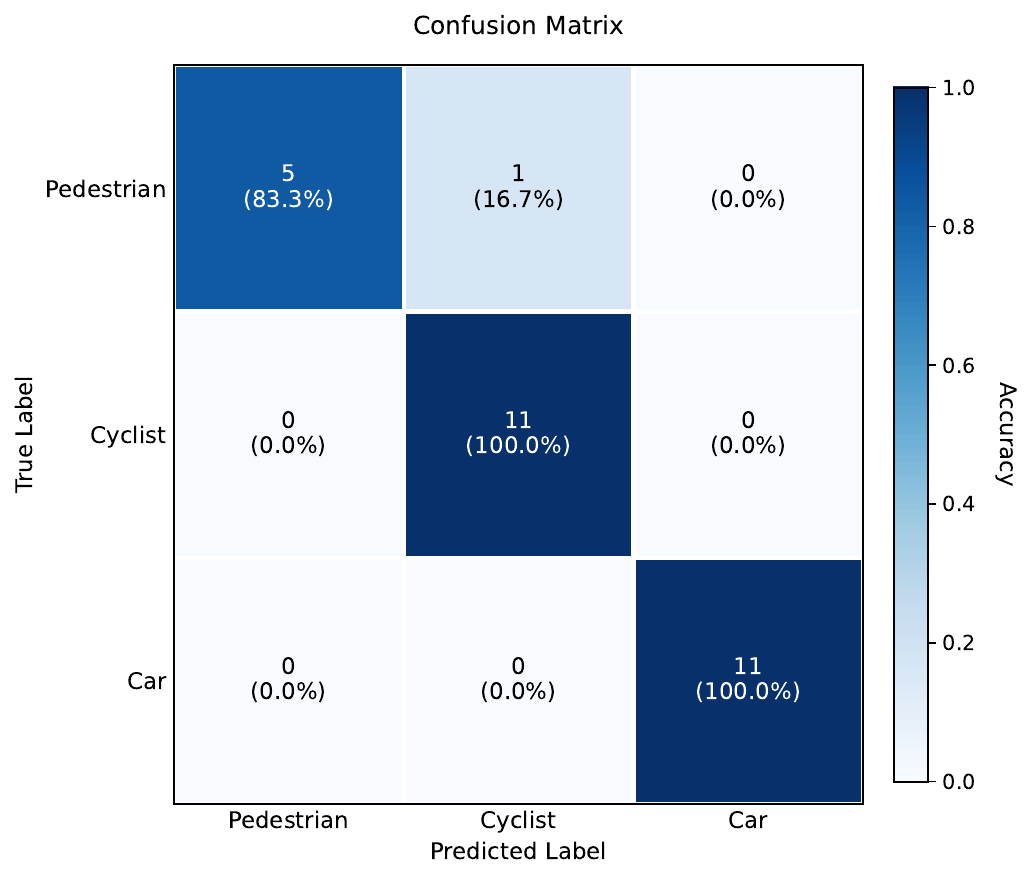}
		\label{fig_confusion_matrix}} 
        \caption{(a) Comparison of models based on the accuracy matrix given $\mathcal{D}_{\rm train}$ data. (b) Confusion matrix of $K_{\rm tar} = 3$ classes on $\mathcal{D}_{\rm test}$ data.}
	\label{fig_accuracy_confusionMatrix} 
\end{figure}

\begin{figure}
    \centering
    \includegraphics[width=\linewidth]{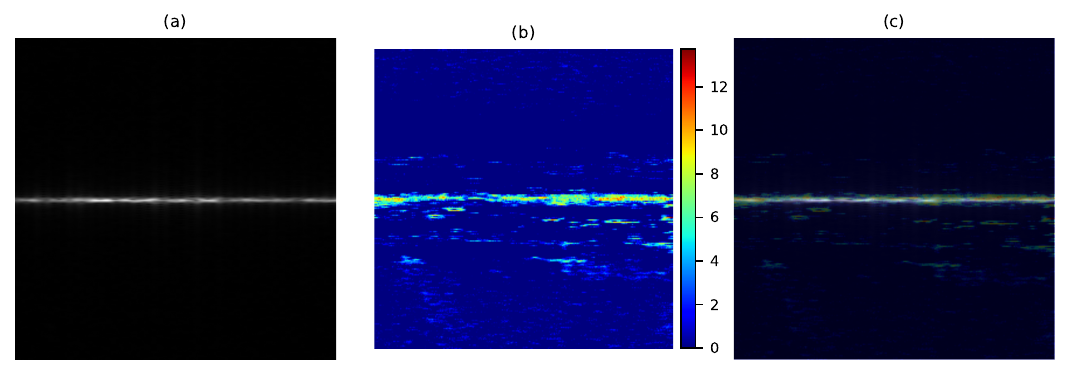}
    \caption{(a) MDS data of car. (b) Attention heatmap. (c) Overlay heatmap.}
    \label{fig_explainable_ai_of_car}
\end{figure}

From Algorithm~\ref{algo_temporal_vit_training}, the numerical accuracy of the models is presented in Fig.~\ref{fig_accuracy_confusionMatrix}a. The T-MDS-ViT converges faster and is more stable compared to the CNN-based methods, i.e., ResNet50 and VGG16, after a few epochs. Although T-MDS-ViT has more trainable parameters than ResNet50 (Table~\ref{tab_model_comparison}), its model size and inference time are more efficient, since the attention mechanism requires fewer computations during both forward and backward passes compared to CNN-based models. Fig.~\ref{fig_accuracy_confusionMatrix}b demonstrates the confusion matrix of each class based on very small $\mathcal{D}_{\rm test}$. This shows that the T-MDS-ViT model achieves good performance, with misclassification occurring only for a pedestrian.
To explain the attention areas of the T-MDS-ViT model on the distribution of the MDS data of the $i$-th class, $\mathbf{S}_i \in \mathbb{C}^{N^{\rm res}_{\rm s}\times N^{\rm res}_{\theta}\times N_{\rm fft}\times N_{\rm fft}}$, we assume class $i$ corresponds to the car, as shown in Fig.~\ref{fig_explainable_ai_of_car}. The figure demonstrates the STFT representation of size $N_{\rm fft}\times N_{\rm fft}$, where the light color indicates the power of the bins obtained by summing over the range–angle dimensions $N^{\rm res}_{\rm s}\times N^{\rm res}_{\theta}$. The attention areas focus on the central line, which is the high-impact feature of the moving car in Fig.~\ref{fig_mds_all}.

\begin{table}[t]
\centering
\caption{Model size and inference time comparison.}
\label{tab_model_comparison}
\begin{tabular}{lccc}
\toprule
\textbf{Model} &
\makecell[tc]{\textbf{Trainable}\\\textbf{Parameters}} &
\makecell[tc]{\textbf{Size}\\\textbf{(MB)}} &
\makecell[tc]{\textbf{Inference}\\\textbf{Time (ms)}} \\
\midrule
Temporal ViT & 107,039,235 & \textbf{480.79} & \textbf{4.320} \\
VGG16        & 138,419,240 & 859.41 & 7.595 \\
ResNet50     & 25,720,104  & 486.14 & 5.075 \\
\bottomrule
\end{tabular}
\end{table}

\section{Conclusion}
In this paper, we propose a framework that applies the temporal model, T-MDS-ViT, to the MDS-based multiclass target classification problem. We demonstrate that the attention mechanism in our model effectively focuses on the high-movement features of each target, thereby enhancing classification performance. The proposed T-MDS-ViT is efficient for real-time applications, achieving a lower inference runtime and requiring fewer processes and backward passes compared to CNN-based methods. This model opens a new pathway for addressing the complexity of real-world MDS data in target detection problems.  
\bibliographystyle{IEEEtran}
\bibliography{References}
\end{document}